\documentclass[%
 aip,
 amsmath,amssymb,
 reprint,%
]{revtex4-1}

\usepackage{graphicx}
\usepackage{dcolumn}
\usepackage{bm}
\usepackage{lineno}

\usepackage[utf8]{inputenc}
\usepackage[T1]{fontenc}
\usepackage{mathptmx}
\usepackage{etoolbox}
\usepackage{xcolor}
\usepackage{soul}

\makeatletter
\def\@email#1#2{%
 \endgroup
 \patchcmd{\titleblock@produce}
  {\frontmatter@RRAPformat}
  {\frontmatter@RRAPformat{\produce@RRAP{*#1\href{mailto:#2}{#2}}}\frontmatter@RRAPformat}
  {}{}
}%
\makeatother
\begin{document}

\preprint{AIP/123-QED}

\title{Two-Stage Lithium Niobate Nonlinear Photonic Circuits for Low-Crosstalk \centerline{and Broadband All Optical Wavelength Conversion}
\hspace*{\fill}
}

\author{Xiaoting LI}
\affiliation{Department of Electrical Engineering, City University of Hong Kong, Kowloon, Hong Kong, China}
\author{Haochuan LI}%
\affiliation{Department of Electrical Engineering, City University of Hong Kong, Kowloon, Hong Kong, China}
\author{Zhaoxi CHEN}
\affiliation{Department of Electrical Engineering, City University of Hong Kong, Kowloon, Hong Kong, China}
\author{Fei MA}
\affiliation{School of Physics, Sun Yat-sen University, Guangzhou 510275, China}
\author{Ke ZHANG}
\affiliation{Department of Electrical Engineering, City University of Hong Kong, Kowloon, Hong Kong, China}
\author{Wenzhao SUN}
\affiliation{City University of Hong Kong (Dongguan), Dongguan, China}
\affiliation{Center of Information and Communication Technology, City University of Hong Kong Shenzhen Research Institute, Shenzhen, China}
\author{Cheng WANG}
\affiliation{Department of Electrical Engineering, City University of Hong Kong, Kowloon, Hong Kong, China}
\affiliation{State Key Laboratory of Terahertz and Millimeter Waves, City University of Hong Kong, Kowloon, Hong Kong, China}
\email{wenzhao.sun@cityu-dg.edu.cn}
\email{cwang257@cityu.edu.hk}

\begin{abstract}
All optical wavelength converters (AOWCs) that can effectively and flexibly switch optical signals between different wavelength channels are essential elements in future optical fiber communications and quantum information systems. A promising strategy for achieving high-performance AOWCs is to leverage strong three-wave mixing processes in second-order nonlinear nanophotonic devices, specifically thin-film periodically poled lithium niobate (TF-PPLN) waveguides. By exploiting the advantages of sub-wavelength light confinement and dispersion engineering compared with their bulk counterparts, TF-PPLN waveguides provide a viable route for realizing highly efficient and broadband wavelength conversion. Nevertheless, most existing approaches rely on a single TF-PPLN device to perform both frequency doubling of the telecom pump and the wavelength conversion process, resulting in significant crosstalk between adjacent signal channels. Here, we address this challenge by demonstrating a two-stage TF-PPLN nonlinear photonic circuit that integrates a second-harmonic generation module, a signal wavelength conversion module, and multiple adiabatic directional coupler-based pump filters, on a single chip. By decoupling the two nonlinear processes and leveraging the high pump-filtering extinction ratio, we achieve low-crosstalk AOWC with a side-channel suppression ratio exceeding 25 dB, substantially surpassing the performance of single-stage devices. Furthermore, our device exhibits an ultra-broad conversion bandwidth of 110 nm and a relatively high conversion efficiency of -15.6 dB, making it an attractive solution for future photonic systems. The two-stage AOWC design shows promise for low-noise phase-sensitive amplification and quantum frequency conversion in future classical and quantum photonic systems.   
\end{abstract}

\maketitle

\section{\label{sec:level1}INTRODUCTION}
The exponential growth of global internet traffic has underscored the significance of wavelength division multiplexing (WDM) technology \cite{2022EuropeanConference2022}, which enables the simultaneous transmission of multiple signal channels over a single optical fiber. As the number of WDM channels increases and the demand for reconfigurable interconnects intensifies, the need for efficient channel conversion and dynamic resource allocation becomes increasingly pressing. To address this challenge, a dynamic traffic allocation mechanism based on frequency conversion at optical intermediate nodes has been proposed for next-generation multi-band optical networks \cite{qianHighconversionefficiencyAllopticalWavelength2017,minamiLowPenaltyBandSwitchableMultiBand2024a}. Traditional optoelectrical wavelength converters are limited to handling frequency conversion of a single wavelength channel with rigid data format and bit rate requirements\cite{yamawakuLowCrosstalk103Channel$times$102006}, and introduce significant latency to the network. In contrast, all-optical wavelength converters (AOWCs), which utilize all-optical methods to convert the frequency of light signals “on the fly”, provide a more flexible and efficient solution to these limitations. Furthermore, AOWCs are also fundamental building blocks for a variety of classical and quantum photonic applications, including quantum frequency conversion \cite{liEfficientLownoiseSinglephotonlevel2016,singhQuantumFrequencyConversion2019}, near-deterministic single-photon source\cite{joshiFrequencyMultiplexingQuasideterministic2018}, nonlinear distortion mitigation\cite{umekiSimultaneousNonlinearityMitigation2016} and up-conversion-based single-photon detectors\cite{maUpconversionSinglephotonDetectors2018,wangQuantumFrequencyConversion2023a}. 

AOWCs can be achieved through various methods. One approach is to utilize the third-order nonlinearity in semiconductor optical amplifiers\cite{summersDesignOperationMonolithically2007,diezFourwaveMixingSemiconductor1997}, which, however, exhibits a limited tuning range constrained by the semiconductor gain bandwidth, as well as elevated noise and significant crosstalk inherent to the amplification process. In contrast, AOWCs based on pure second ($\chi$$^{(2)}$) or third ($\chi$$^{(3)}$) order optical nonlinear processes are more appealing due to their simplicity in theoretical modeling and minimal addition of noise. For instance, $\chi$$^{(3)}$-based AOWC has been demonstrated in platforms such as silicon and silicon nitride, achieving the simultaneous conversion of multiple wavelength bands with minimal excess noise. However, the relatively weak $\chi$$^{(3)}$ process usually necessitates long highly nonlinear optical fibers\cite{qianHighconversionefficiencyAllopticalWavelength2017} or waveguides\cite{lacavaUltraCompactAmorphousSilicon2016,leeDemonstrationBroadbandWavelength2009,riemensbergerPhotonicIntegratedContinuoustravellingwave2022} to achieve a reasonable frequency conversion efficiency. The conversion efficiency can be substantially enhanced using optical resonators, at the cost of compromised signal bandwidths\cite{liEfficientLownoiseSinglephotonlevel2016,absilWavelengthConversionGaAs2000,haishengrongSiliconBasedChipscale2008}.

In comparison, $\chi$$^{(2)}$-based AOWC could potentially achieve much higher conversion efficiencies by taking advantage of the leading nonlinear optical effect\cite{kazamaOver30dBGain1dB2021b}. Additionally, $\chi$$^{(2)}$-based AOWCs have shown relatively smaller crosstalk with increasing number of input signal channels\cite{szaboNumericalComparisonWDM2014} and broader tuning range compared with fiber-based $\chi$$^{(3)}$ counterparts\cite{yamawakuLowCrosstalk103Channel$times$102006,kazamaLowParametricCrosstalkPhaseSensitiveAmplifier2017c}. Traditionally, $\chi$$^{(2)}$-based AOWCs typically rely on bulk or weakly-guided periodically poled lithium niobate (PPLN) waveguides with limited nonlinear interaction strengths, necessitating device lengths of several centimeters. This not only increases the system size and cost, but also limits the phase-matching bandwidth of the nonlinear processes. Furthermore, an efficient and low-crosstalk AOWC system requires several nonlinear optical modules with high performance consistency, which are challenging due to the limited scalability of the traditional lithium niobate platform, necessitating separate fabrication and implementation of these modules. 

The emergence of the thin-film periodically poled lithium niobate (TF-PPLN) platform offers a promising next-generation solution for AOWCs with high integration level, broad operation bandwidth, and high conversion efficiency. Thanks to its sub-wavelength light confinement and capability for dispersion engineering, TF-PPLN waveguides exhibit dramatically higher normalized conversion efficiency \cite{wangUltrahighefficiencyWavelengthConversion2018a,chenAdaptedPolingBreak2024,heEfficientSecondHarmonicGeneration2024} and broader bandwidth\cite{jankowskiUltrabroadbandNonlinearOptics2020a} in nonlinear wavelength conversion processes. More importantly, the excellent scalability of the TF-PPLN platform now enables the integration of multiple linear and nonlinear devices on a single chip through wafer-scale fabrication\cite{liAdvancingLargescaleThinfilm2024}, enabling a variety of functional nonlinear photonic integrated circuits. These include flat-top optical frequency comb generator\cite{stokowskiIntegratedFrequencymodulatedOptical2024b}, octave-spanning optical parametric oscillators\cite{ledezmaOctavespanningTunableInfrared2023d,hwangMidinfraredSpectroscopyBroadly2023f}, and integrated Pockels laser\cite{liIntegratedPockelsLaser2022e}. However, existing TF-PPLN AOWCs have been limited to demonstrations based on a single nonlinear waveguide, where the two nonlinear processes, i.e., pump second-harmonic generation (SHG) and difference-frequency generation (DFG)-based signal conversion processes, occur simultaneously\cite{weiEfficientBroadbandAllOptical2024,weiAllopticalWavelengthConversion2022a}. As a result, input signal light also mixes with the telecom pump light through undesired sum-frequency generation (SFG) processes, leading to substantial crosstalk between neighboring signal channels. In principle, crosstalk-free AOWC can be achieved by directly pumping at the second-harmonic (SH)  frequency, which however would rely on costly near-visible laser sources.

In this article, we propose and demonstrate a two-stage AOWC that achieves minimal signal crosstalk, broad operation bandwidth, and compact system size. Our AOWC design consists of two TF-PPLN waveguides and a wavelength multiplexer comprising three cascaded adiabatic directional couplers (ADCs), monolithically integrated on a single nonlinear photonic chip. Experimental results showcase an efficient AOWC, which boasts a relatively high conversion efficiency of -15.6 dB, a broad operation bandwidth of 110 nm, and a 25 dB suppression in signal crosstalk compared to single TF-PPLN-based AOWCs. These findings demonstrate the feasibility of TF-PPLN devices in achieving high-efficiency and low-crosstalk AOWC, which holds promise for future applications in high-performance chip-integrated optical parametric amplifiers and phase-sensitive amplifiers.
\section{\label{sec:level2}DESIGN AND FABRICATION}
\begin{figure*}[htb]
\includegraphics{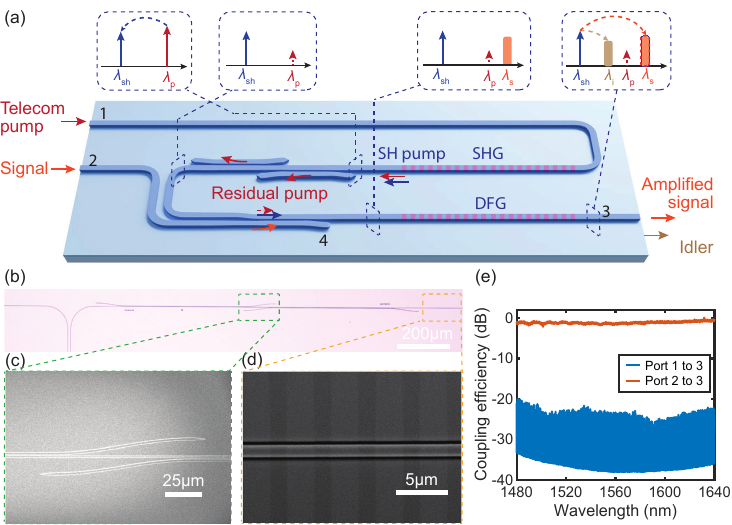}
\caption{\label{fig:one}(a) Schematic of the proposed nonlinear photonic circuit for all-optical wavelength conversion. Telecom pump light is first converted to its second-harmonic (SH) frequency via a first periodically poled lithium niobate (PPLN) module. After filtering out the residual telecom pump, the SH pump mixes with the signal light to generate the converted signal light (i.e., the idler light). (b) Optical microscope images of the fabricated device. (c-d) Scanning electron microscope images of the adiabatic directional coupler (ADC, c) and the PPLN region (d).(e) Measured coupling efficiency of our fabricated ADC from port 1 to 3 (blue), and from port 2 to 3 (red) in the telecom band.}
\end{figure*}
Figure~\ref{fig:one}(a) schematically illustrates the proposed chip-scale AOWC circuit consisting of three main modules: (i) a pump frequency doubling module that performs SHG of telecom-band pump light at $\omega$$_p$ via a first PPLN waveguide to 2$\omega$$_p$, which serves as the SH pump light for subsequent processes; (ii) visible-telecom wavelength multiplexers that filter out the residual telecom pump light and combine the SH pump with signal light $\omega$$_s$, which is designed and implemented using a sequence of three adiabatic directional couplers (ADCs); (iii) a DFG module that converts the input signal light to idler light at $\omega$$_i$$=$$2$$\omega$$_p$-$\omega$$_s$. The circuit features four ports, among which port 1, 2, and 3 are the telecom-pump input, signal input, and the amplified signal output ports, respectively. A tap channel, i.e., port 4, is designed to evaluate the coupling ratio of the adiabatic taper. 

Figure~\ref{fig:one}(b) presents an optical microscopy image of the fabricated device, with zoom-in view of the ADC structure detailed in Fig.~\ref{fig:one}(c). The ADCs are engineered with adiabatically tapered width profiles, as described in Ref. \cite{sunAdiabaticityCriterionShortest2009}, while the waveguide gap is kept at a constant value of 1.1 $\mu$m. This design allows the generated SH pump to pass through with minimal disturbance while effectively filtering out the residual telecom-band pump light and coupling in the signal light simultaneously. The total coupling length of each ADC is chosen to be 450 $\mu$m, optimized to facilitate a high coupling efficiency while minimizing the overall device footprint. According to our simulation results, the coupling efficiency of telecom light for single ADC can reach 96$\%$ within a working bandwidth of 100 nm, while the coupling loss for SH light is less than 0.02 dB. 

Figure~\ref{fig:one}(d) shows a zoom-in scanning electron microscope (SEM) image of the poling region, highlighting high poling fidelity with a duty cycle of approximately 40$\%$. The non-ideal poling duty cycle is attributed to the suboptimal poling conditions during the domain inversion process, and can be improved by better poling waveform design and calibration. The PPLN waveguides feature a 600-nm total thickness, a 220-nm etching depth, and a 60-degree sidewall angle with silica cladding. We use an identical waveguide width of $\sim950$ $ \rm{nm}$ and poling period of 4.3 $\rm{\mu}\rm{m}$ for both PPLN waveguides to ensure consistent quasi-phase-matching (QPM) peak wavelength for both SHG and DFG processes. This design yields a theoretical absolute conversion efficiency of 4,500$\%$W$^{-1}$cm$^{-2}$. The lengths of the two PPLN waveguides are both 6 $\rm{mm}$.

The overall conversion efficiency of the AOWC, $\eta_{AOWC}$, depends on the efficiencies of both nonlinear optical processes and the coupling and propagation losses in the intermediate section, as expressed in Eq. \ref{eq:one}: 
\begin{eqnarray}
\eta_{AOWC}&&=\frac{P_{idler}} {P_{signal}}=\frac{P_{idler}}{P_{signal}\cdot{P_{SH,2}}}\cdot{P_{SH,2}}\nonumber\\
&&=\eta_{DFG}\cdot(\kappa_{ADC}\cdot{P_{SH,1}})\nonumber\\
&&=\eta_{DFG}\cdot\kappa_{ADC}\cdot\eta_{SHG}\cdot(\alpha\cdot{{P_{pump}}^2}).
\label{eq:one}
\end{eqnarray}
, where $\eta$$_{SHG}$ and $\eta$$_{DFG}$  represent the normalized conversion efficiencies for SHG and DFG processes in PPLN 1 and PPLN 2, respectively; $\kappa$$_{ADC}$ is the SH pump coupling loss in the three cascaded ADC devices; $\alpha$ is the on-chip telecom pump propagation loss before entering the PPLN 1; $P$$_{pump}$ and $P$$_{signal}$ are the input telecom pump and signal power; $P_{idler}$ is the converted idler light power; $P$$_{SH,1}$ and $P$$_{SH,2}$ stand for the SH pump light generated from PPLN 1 and that entering the PPLN 2, respectively, which differ by $\kappa$$_{ADC}$. All powers here correspond to on-chip powers excluding fiber-chip coupling losses. Overall, the conversion effciency of the two-stage AOWC goes quadratically with the input telecom pump power and is independent from the input signal power.

We fabricate our device using a wafer-scale fabrication and poling process established in our previous work \cite{liAdvancingLargescaleThinfilm2024}. Specifically, a 600 nm thick magnesium-doped lithium niobate-on-insulator (LNOI) wafer (NANOLN) is first patterned with poling electrodes in selected regions for SHG and DFG processes via ASML ultraviolet (UV) stepper photolithography, followed by electron-beam evaporation of chromium/gold and a standard lift-off process. An automated poling process applies high-voltage electrical pulses to each set of poling electrodes in sequence utilizing the same methodology as described in \cite{liAdvancingLargescaleThinfilm2024}, ensuring consistent and precise poling. After removing all poling electrodes, the optical waveguides are formed using a second aligned stepper lithography step, followed by an inductively coupled plasma reactive ion etching (ICP-RIE) process. The surface is then cleaned and protected with an 800 nm thick silica cladding. Finally, we cleave the edges of the chip for efficient end-fire coupling using a fiber array (FA).

\section{\label{sec:level2}RESULTS AND DISCUSSION}
\subsection{\label{sec:level2}Linear Performance Characterization}
\begin{figure*}[t]
\includegraphics{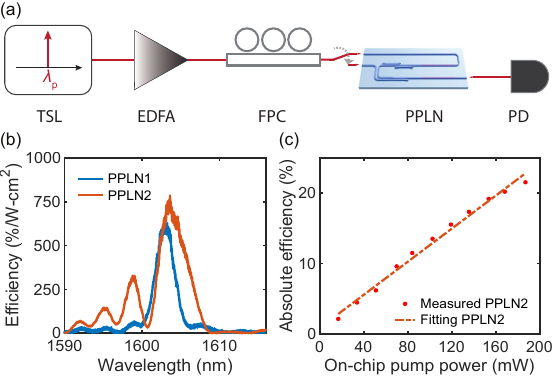}
\caption{\label{fig:two} (a) Schematic of the second-harmonic generation (SHG) characterization setup. (b) Measured normalized SHG conversion efficiency versus pump wavelength for the two TF-PPLN devices. (c) Measured and fitted absolute SHG conversion efficiency versus on-chip pump power. TLS, tunable laser source; EDFA, erbium-doped fiber amplifier; PC, polarization controller; PD, photodetector.}
\end{figure*}
To comprehensively evaluate the performance of individual modules within the nonlinear circuit, we first conduct a series of meticulous linear optical characterizations by measuring the transmission loss between channels 1 to 3, 1 to 4, 2 to 3, and 2 to 4 at both the QPM wavelength of 1603 nm and its corresponding SH wavelength of 801.5 nm. For telecom band measurements, we inject light from a tunable continuous-wave (CW) laser (Santec TSL-710) into the devices using a lensed fiber. A fiber polarization controller (FPC) is used to excite fundamental transverse-electric ($TE$$_{00}$) mode, which aligns with the largest $d$$_{33}$ nonlinear tensor component of lithium niobate (LN) for later nonlinear experiments. Figure~\ref{fig:one}(e) shows the measured transmission coefficients of the ADC in the telecom band when injecting light from port 1 (blue) and port 2 (red), and measuring the output from port 3. The measured extinction ratio (ER) of 24.5 dB at the QPM wavelength of 1603 nm indicates a strong suppression of the residual telecom-band pump light after passing through the three ADCs, while maintaining minimal loss for coupling in the signal light from port 2. The fluctuation in the measured crosstalk (port 1 to 3) mainly stems from the noise limit of our photodetection system. Thanks to the adiabatic coupler design, high coupling ER can be achieved across the entire measured wavelength range between 1480 nm and 1640 nm, ensuring robust system performance against potential QPM wavelength shift caused by fabrication deviation.

\begin{figure*}[ht]
\includegraphics{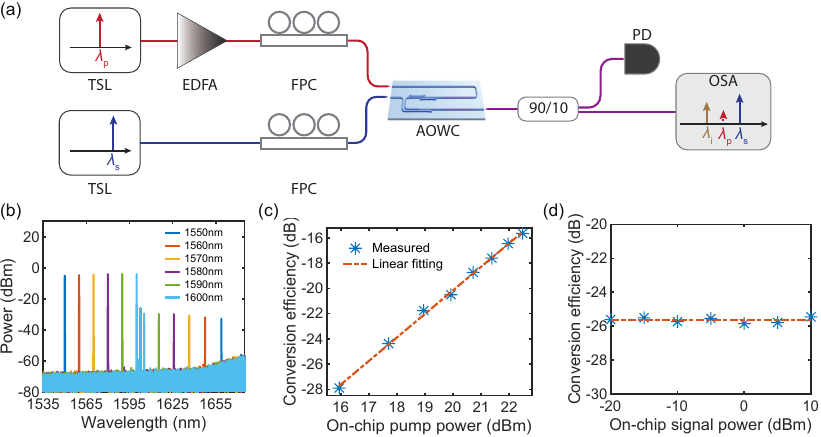}
\caption{\label{fig:three}(a) Schematic of the two-stage all-optical wavelength converter (AOWC) characterization setup. (b)   Measured AOWC spectra at various input signal wavelengths with the on-chip pump power fixed at 17.2 dBm. (c) Measured AOWC efficiency versus on-chip pump power with the on-chip signal power fixed at -15 dBm. (d) Measured conversion   efficiency versus the on-chip signal power with the on-chip pump power fixed at 17.2 dBm.}
\end{figure*}

We then input a near-visible light source (M-Squared SOLSTIS PI) at the SH wavelength of 801.5 nm into port 2 and compare the output power difference between ports 3 and 4. The results reveal a transmission difference of -7 dB, indicating more than 80$\%$ of the generated SH pump could remain in the original waveguide through one ADC structure. This suggests that our ADC design can efficiently multiplex visible and telecom-band signals. Further characterizations of
other combinations of input and output ports, as well as comparison
with reference devices on the same chip, yield estimated
waveguide propagation losses of 2.2 dB/cm in the near-visible
band and 1 dB/cm in the telecom band (see Supplementary Information for details). The propagation loss can be further reduced by optimizing the fabrication process, including the dry etching and post-etching annealing techniques \cite{zhuTwentynineMillionIntrinsic2024}, and adopting waveguide designs with wider widths or shallower etch depths \cite{zhaoShallowetchedThinfilmLithium2020h}.The edge coupling efficiency is estimated to be 4.5 dB/facet and 6 dB/facet for telecom and visible bands, respectively.

\subsection{\label{sec:level2}Second Harmonic Generation (SHG) Characterization}

We further evaluate the nonlinear wavelength conversion performances of the two TF-PPLN modules by conducting a thorough examination of their QPM spectra and conversion efficiencies using the experimental setup depicted in Fig.~\ref{fig:two}(a). We characterize the two PPLN devices by injecting telecom pump light to ports 1 and 2, respectively, and collect the output SH signal from port 3 using another lensed fiber connected to a visible photodetector (PD, Newport 1801). Figure~\ref{fig:two}(b) shows the measured small-signal SHG efficiencies of the two fabricated PPLN devices as functions of pump wavelength, indicating nearly aligned peak QPM wavelengths of 1603 nm and 1603.7 nm. The normalized conversion efficiencies of the two TF-PPLN devices are approximately 625$\%$W$^{-1}$cm$^{-2}$ and 750$\%$W$^{-1}$cm$^{-2}$, which are estimated by considering the propagation loss, edge coupling loss, and the ADC coupling ratios at both visible and telecom bands, using the linear measurement results obtained earlier. The measured conversion efficiencies are lower than the theoretical value in this batch of device fabrication likely due to unsatisfactory poling depth and film thickness variation, and can be substantially improved by further optimizing the poling waveform \cite{changThinFilmWavelength2016d,zhaoOpticalDiagnosticMethods2019b,nagySituTemporalPeriodic2020}, and introducing segmented microheaters for QPM fine tuning \cite{liAdvancingLargescaleThinfilm2024}.

To achieve an efficient AOWC, the TF-PPLN devices are required to operate in the high-power regime with substantial absolute power conversion. To characterize the device performance in this regime, we further amplify the input pump light using an erbium-doped fiber amplifier (EDFA, Amonics, AEDFA-L-30-B-FA). The measured absolute SHG efficiency reaches approximately 20$\%$ under an on-chip input power of 190 mW [Fig.~\ref{fig:two}(c)].

\subsection{\label{sec:level2}Two-stage AOWC Performance Characterization}

We next characterize the conversion efficiency and gain bandwidth of the full two-stage AOWC using the experimental setup depicted in Fig.~\ref{fig:three}(a). In this experiment, the telecom pump laser is fixed at the peak QPM wavelength of 1603 nm, whereas another tunable CW laser (Santec TSL-550, 1500-1630 nm) at a nearby wavelength is used as signal light source for the AOWC process. The two optical signals are input to ports 1 and 2 of the two-stage AOWC chip via an FA. At the chip output end of port 3, a 90/10 splitter is utilized to simultaneously observe the SH power using the visible photodetector and the output spectra using an optical spectrum analyzer (OSA). The SH power at the Si-PD serves as an indicator for optimized polarization and coupling state during the measurements. 

Firstly, we characterize the operation bandwidth by sweeping the signal wavelength while fixing the on-chip pump power at 17.2 dBm. As shown in Fig.~\ref{fig:three}(b), effective conversion can be consistently achieved at signal wavelengths from 1550 nm to 1600 nm, resulting in idler light at 1610 nm to 1660 nm. The conversion efficiency drops by 3 dB at an idler wavelength of 1660 nm due to dispersion-induced gradual phase-matching walk-off. Due to the reciprocity between signal and idler wavelengths, we estimate the full 3-dB bandwidth of the AOWC to be 110 nm, i.e., from 1550 nm to 1660 nm. 

We further increase the pump power to achieve higher absolute conversion efficiency, as shown in Fig.~\ref{fig:three}(c). The measured conversion efficiency clearly increases with increasing on-chip pump power, with a fitted slope of 1.9 in double-logarithmic scale, which aligns with the expected quadratic input-output power relationship according to Eq. \ref{eq:one}. The maximum measured conversion efficiency reaches -15.6 dB under an on-chip pump power of 22.8 dBm, currently limited primarily by the non-ideal poling and ADC performance in the current batch of devices. Assuming a higher normalized conversion efficiency of 3,800$\%$W$^{-1}$cm$^{-2}$ achieved in our earlier work \cite{liAdvancingLargescaleThinfilm2024}, and a reduced visible coupling loss of 0.07 dB per ADC, the converison efficiency could potentially be enhanced to above 10 dB under an on-chip pump power of 250 mW \cite{chenHighgainOpticalParametric2024}. This would be highly relevant for future applications in optical parametric amplifiers and ultra-low noise phase-sensitive amplifiers. On the other hand, the conversion efficiency is insensitive to input signal power, as illustrated in Fig.~\ref{fig:three}(d), with an average value of -25.6 dB (dashed line) for a fixed on-chip pump power of 17.2 dBm.

\begin{figure*}[ht]
\includegraphics{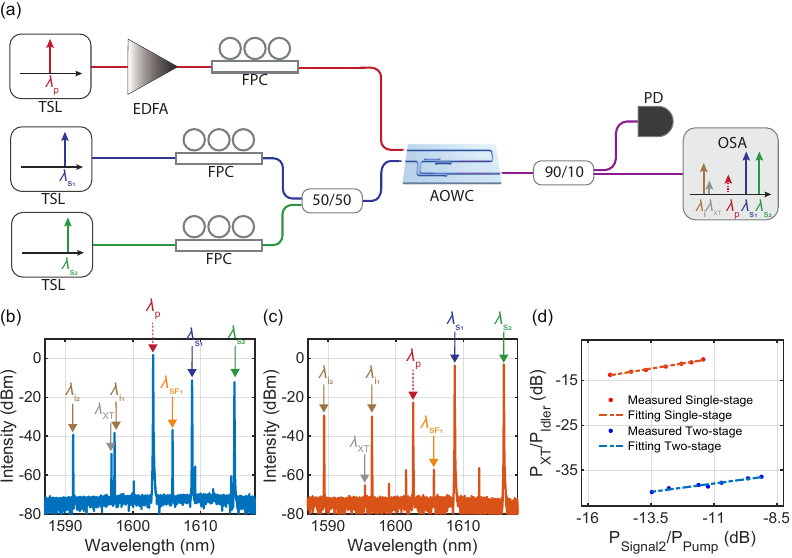}
\caption{\label{fig:four}(a) Schematic of the experimental set-up for crosstalk characterization. (b-c) Optical spectra of the single-stage (b) and two-stage (c) AOWCs when injecting two signals simultaneously. (d) Dependence of the generated crosstalk on input signal powers.}
\end{figure*}

\subsection{\label{sec:level2}Low-crosstalk AOWC Characterization}
As previously discussed, signal crosstalk due to nonlinear interactions between different signal channels can significantly degrade the system performance in DWDM applications. In the case of single-stage $\chi$$^{(2)}$-based AOWC devices, the pump light at $\omega$$_{0}$ not only is frequency doubled through the desired SHG process, but also inevitably mixes with input signal light $\omega$$_{s_1}$ leading to an SFG signal at $\omega$$_{SF}$ = $\omega$$_{0}$ + $\omega$$_{s_1}$. The generated $\omega$$_{SF}$ light could further interact with neighboring signals, e.g. at $\omega$$_{s_2}$, via an undesired DFG process. This creates a crosstalk signal at $\omega$$_{XT}$$^{'}$ = $\omega$$_{SF}$ - $\omega$$_{s_2}$ = $\omega$$_0$ + $\omega$$_{s_1}$ - $\omega$$_{s_2}$, which could be spectrally close to the target idler frequency $\omega_{idler}$ = 2$\omega_0$ - $\omega_{s_1}$ if $\omega_{s_1}$ - $\omega_{s_2}$ = $\omega_0$ - $\omega_{s_1}$. In contrast, our approach minimizes the direct nonlinear interaction between the telecom pump and signal by effectively filtering out the residual telecom pump after the SHG module, thereby reducing the crosstalk and preserving the quality of the converted signal.
To evaluate the crosstalk reduction performance of our device in comparison with existing single-stage AOWC schemes, we introduce a second telecom signal laser using the setup depicted in Fig.~\ref{fig:four}(a). The two signal paths are combined using an off-chip 50/50 fiber coupler before injecting into the chip through the FA. We keep an identical on-chip power level for the two signal lasers during the measurement. Figure~\ref{fig:four}(b-c) shows the measured output spectra of a single-stage and the proposed two-stage AOWCs, respectively. To make this contrast experiment more convincing, we evaluate the performance of single-stage AOWC (as reference) by injecting both pump and signal light into port 2 of the two-stage device and bypassing the first PPLN, such that the same PPLN 2 is used for AOWC in both experiments. As shown in Fig.~\ref{fig:four}(b), the single-stage AOWC generates a significant crosstalk signal $\lambda$$_{XT}$ near the target idler light at $\lambda$$_{i_1}$, featuring a signal-to-crosstalk ratio of less than 11 dB. In contrast, as shown in Fig.~\ref{fig:four}(c), our two-stage AOWC exhibits strongly suppressed crosstalk with a signal-to-crosstalk ratio of ~40 dB, thanks to the de-coupled pump-SHG and signal-conversion processes. The significant suppression of the undesired SFG process can also be observed from the SFG signals in the measured spectra (labeled as $\lambda$$_{{SF}_{1}}$), which is an artifact half-harmonic signal from the OSA but qualitatively indicates the difference in the SFG signals from the two devices.

To further quantitatively investigate the crosstalk of the two AOWC architectures, we measure the power ratio between crosstalk and idler at different input signal power, as illustrated in Fig.~\ref{fig:four}(d). Since the crosstalk is primarily generated through the cascaded SFG and DFG nonlinear processes, which can be expressed as
\begin{equation}
P_{XT}
\propto{\eta_{DFG}}\cdot{P_{s_2}}\cdot{P_{SF_1}}\propto{\eta_{DFG}}\cdot{P_{s_2}}\cdot({\eta_{SFG}}\cdot{P_{pump}}\cdot{P_{s_1}})
\label{eq:two}
\end{equation}
, where $P_{pump}$, $P_{s_1}$, $P_{s_2}$ represent the powers of input telecom pump, signal 1, and signal 2, respectively; $\eta_{DFG}$ and $\eta_{SFG}$ denote the normalized nonlinear conversion efficiencies of the DFG and SFG processes. For simplicity, here we ignore the propagation and ADC coupling losses as these linear terms do not affect the qualitative trend analysis. On the other hand, the target idler signal $P_{idler}$ is proportional to ${P_{pump}}^2\cdot{P_{s_1}}$ following Eq. \ref{eq:one}. Therefore, the crosstalk-to-signal ratio can be estimated as:
\begin{equation}
\frac{P_{XT}} {P_{idler}}
\propto\frac{{P_{pump}}\cdot{P_{s_1}}\cdot{P_{s_2}}}{{P_{pump}}^2\cdot{P_{s_1}}}\\
=\frac{{P_{s_2}}}{P_{pump}}
\label{eq:three}
\end{equation}
This relationship indicates that the crosstalk-to-idler ratio increases linearly with the interfering signal power $P_{s_2}$ for a fixed telecom pump power, leading to a theoretically expected slope of 1 in a double-logarithmic scale in Fig.~\ref{fig:four}(d). The fitted slopes of actual experimental results are approximately 0.8 for the two-stage AOWC and 0.9 for the single-stage AOWC, respectively. The more substantial deviation in the two-stage case may be attributed to measurement uncertainty from the exceptionally low crosstalk level of $^{\sim}$-70 dBm approaching the noise floor. Importantly, the two-stage AOWC device (blue) shows consistent crosstalk suppression ratio of 25 dB compared with the single-stage reference (red) at various signal power levels, which is crucial for practical applications where densely packed wavelength channels are used simultaneously. The crosstalk situations in multi-channel input (> 2) scenarios are discussed in more details in the Supplementary Information.

\section{\label{sec:level2}CONCLUSION}
In conclusion, we present a two-stage chip-integrated AOWC, which exhibits a broad working bandwidth of 110 nm, a relatively high conversion efficiency of -15.6 dB, and significant potential for large-scale integration. The two-stage AOWC architecture leads to a remarkable 25 dB reduction in crosstalk compared to AOWC based on a single TF-PPLN device. The absolute conversion efficiency can be further improved by optimizing the ADC and PPLN performance, as well as increasing the edge-coupling efficiency between the lensed fiber and the chip. By introducing a copier module and thermal-optic phase shifters \cite{maederHighbandwidthThermoopticPhase2022}, the realization of a fully integrated, ultra-low-noise phase-sensitive amplifier becomes potentially feasible based on the fabrication method and device architecture in this work. The demonstrated two-stage AOWC structure has the potential to play a crucial role in various applications, including high-speed optical communication systems, quantum information processing, and spectroscopy.

\section{SUPPLEMENTARY MATERIAL}
See the supplementary material for additional information.

\begin{acknowledgments}
The authors would like to acknowledge the Research Grants Council, University Grants Committee (N$\_$CityU113/20, CityU 11215024, STG3/E-704/23-N; CityU 11204523); Innovation Program for Quantum Science and Technology (No. 2021ZD0301500); Croucher Foundation (9509005); City University of Hong Kong (9610682); University Grants Committee (PF18-17958), for funding this work. The authors would like to thank Mr. Hanke FENG for his valuable discussions and assistance in the fabrication process.
\end{acknowledgments}
\section*{AUTHOR DECLARATIONS}

\section*{\label{sec:level2}Conflict of Interest}
The authors have no conflicts to disclose.

\section*{\label{sec:level2}Author Contributions}
X.L. and H.L. contributed equally to this work.

Xiaoting Li: Conceptualization (equal); Investigation (equal); Writing – original draft (lead). Haochuan Li: Investigation (equal); Writing – original draft (supporting). Zhaoxi Chen: Investigation (supporting). Fei Ma: Investigation (supporting). Ke Zhang: Investigation (supporting). Wenzhao Sun: Conceptualization (supporting); Investigation (supporting); Supervision (equal); Writing – review editing (equal). Cheng Wang: Conceptualization (equal); Funding acquisition (lead); Investigation (supporting); Supervision (equal); Writing – review editing (equal).

\section*{Data Availability Statement}
Data underlying the results presented in this paper are not publicly available at this time but may be obtained from the corresponding author upon reasonable request.

\section*{references}
\nocite{*}
\bibliography{aowctry2}

\begin{thebibliography}{39}%
\makeatletter
\providecommand \@ifxundefined [1]{%
 \@ifx{#1\undefined}
}%
\providecommand \@ifnum [1]{%
 \ifnum #1\expandafter \@firstoftwo
 \else \expandafter \@secondoftwo
 \fi
}%
\providecommand \@ifx [1]{%
 \ifx #1\expandafter \@firstoftwo
 \else \expandafter \@secondoftwo
 \fi
}%
\providecommand \natexlab [1]{#1}%
\providecommand \enquote  [1]{``#1''}%
\providecommand \bibnamefont  [1]{#1}%
\providecommand \bibfnamefont [1]{#1}%
\providecommand \citenamefont [1]{#1}%
\providecommand \href@noop [0]{\@secondoftwo}%
\providecommand \href [0]{\begingroup \@sanitize@url \@href}%
\providecommand \@href[1]{\@@startlink{#1}\@@href}%
\providecommand \@@href[1]{\endgroup#1\@@endlink}%
\providecommand \@sanitize@url [0]{\catcode `\\12\catcode `\$12\catcode `\&12\catcode `\#12\catcode `\^12\catcode `\_12\catcode `\%12\relax}%
\providecommand \@@startlink[1]{}%
\providecommand \@@endlink[0]{}%
\providecommand \url  [0]{\begingroup\@sanitize@url \@url }%
\providecommand \@url [1]{\endgroup\@href {#1}{\urlprefix }}%
\providecommand \urlprefix  [0]{URL }%
\providecommand \Eprint [0]{\href }%
\providecommand \doibase [0]{http://dx.doi.org/}%
\providecommand \selectlanguage [0]{\@gobble}%
\providecommand \bibinfo  [0]{\@secondoftwo}%
\providecommand \bibfield  [0]{\@secondoftwo}%
\providecommand \translation [1]{[#1]}%
\providecommand \BibitemOpen [0]{}%
\providecommand \bibitemStop [0]{}%
\providecommand \bibitemNoStop [0]{.\EOS\space}%
\providecommand \EOS [0]{\spacefactor3000\relax}%
\providecommand \BibitemShut  [1]{\csname bibitem#1\endcsname}%
\let\auto@bib@innerbib\@empty
\bibitem [{202(2022)}]{2022EuropeanConference2022}%
  \BibitemOpen
  \href@noop {} {\emph {\bibinfo {title} {2022 {{European Conference}} on {{Optical Communication}} ({{ECOC}}): 18-22 {{Sept}}. 2022}}}\ (\bibinfo  {publisher} {IEEE},\ \bibinfo {address} {Piscataway, NJ},\ \bibinfo {year} {2022})\BibitemShut {NoStop}%
\bibitem [{\citenamefont {Qian}\ \emph {et~al.}(2017)\citenamefont {Qian}, \citenamefont {Chen}, \citenamefont {Chen},\ and\ \citenamefont {Gao}}]{qianHighconversionefficiencyAllopticalWavelength2017}%
  \BibitemOpen
  \bibfield  {author} {\bibinfo {author} {\bibfnamefont {J.}~\bibnamefont {Qian}}, \bibinfo {author} {\bibfnamefont {B.}~\bibnamefont {Chen}}, \bibinfo {author} {\bibfnamefont {W.}~\bibnamefont {Chen}}, \ and\ \bibinfo {author} {\bibfnamefont {M.}~\bibnamefont {Gao}},\ }\bibfield  {title} {\enquote {\bibinfo {title} {High-conversion-efficiency all-optical wavelength converter based on cascaded highly nonlinear fibers},}\ }\href {\doibase 10.1364/AO.56.005871} {\bibfield  {journal} {\bibinfo  {journal} {Applied Optics}\ }\textbf {\bibinfo {volume} {56}},\ \bibinfo {pages} {5871} (\bibinfo {year} {2017})}\BibitemShut {NoStop}%
\bibitem [{\citenamefont {Minami}\ \emph {et~al.}(2024)\citenamefont {Minami}, \citenamefont {Hirose}, \citenamefont {Fukatani}, \citenamefont {Nakagawa}, \citenamefont {Seki}, \citenamefont {Shimizu}, \citenamefont {Kobayashi}, \citenamefont {Kazama}, \citenamefont {Enbutsu}, \citenamefont {Umeki},\ and\ \citenamefont {Kuwahara}}]{minamiLowPenaltyBandSwitchableMultiBand2024a}%
  \BibitemOpen
  \bibfield  {author} {\bibinfo {author} {\bibfnamefont {H.}~\bibnamefont {Minami}}, \bibinfo {author} {\bibfnamefont {K.}~\bibnamefont {Hirose}}, \bibinfo {author} {\bibfnamefont {T.}~\bibnamefont {Fukatani}}, \bibinfo {author} {\bibfnamefont {M.}~\bibnamefont {Nakagawa}}, \bibinfo {author} {\bibfnamefont {T.}~\bibnamefont {Seki}}, \bibinfo {author} {\bibfnamefont {S.}~\bibnamefont {Shimizu}}, \bibinfo {author} {\bibfnamefont {T.}~\bibnamefont {Kobayashi}}, \bibinfo {author} {\bibfnamefont {T.}~\bibnamefont {Kazama}}, \bibinfo {author} {\bibfnamefont {K.}~\bibnamefont {Enbutsu}}, \bibinfo {author} {\bibfnamefont {T.}~\bibnamefont {Umeki}}, \ and\ \bibinfo {author} {\bibfnamefont {T.}~\bibnamefont {Kuwahara}},\ }\bibfield  {title} {\enquote {\bibinfo {title} {Low-{{Penalty Band-Switchable Multi-Band Optical Cross-Connect Using PPLN-Based Inter-Band Wavelength Converters}}},}\ }\href {\doibase 10.1109/JLT.2023.3302475} {\bibfield  {journal} {\bibinfo  {journal} {Journal of Lightwave Technology}\ }\textbf
  {\bibinfo {volume} {42}},\ \bibinfo {pages} {1242--1249} (\bibinfo {year} {2024})}\BibitemShut {NoStop}%
\bibitem [{\citenamefont {Yamawaku}\ \emph {et~al.}(2006)\citenamefont {Yamawaku}, \citenamefont {Takara}, \citenamefont {Ohara}, \citenamefont {Takada}, \citenamefont {Morioka}, \citenamefont {Tadanaga}, \citenamefont {Miyazawa},\ and\ \citenamefont {Asobe}}]{yamawakuLowCrosstalk103Channel$times$102006}%
  \BibitemOpen
  \bibfield  {author} {\bibinfo {author} {\bibfnamefont {J.}~\bibnamefont {Yamawaku}}, \bibinfo {author} {\bibfnamefont {H.}~\bibnamefont {Takara}}, \bibinfo {author} {\bibfnamefont {T.}~\bibnamefont {Ohara}}, \bibinfo {author} {\bibfnamefont {A.}~\bibnamefont {Takada}}, \bibinfo {author} {\bibfnamefont {T.}~\bibnamefont {Morioka}}, \bibinfo {author} {\bibfnamefont {O.}~\bibnamefont {Tadanaga}}, \bibinfo {author} {\bibfnamefont {H.}~\bibnamefont {Miyazawa}}, \ and\ \bibinfo {author} {\bibfnamefont {M.}~\bibnamefont {Asobe}},\ }\bibfield  {title} {\enquote {\bibinfo {title} {Low-{{Crosstalk}} 103 {{Channel}}\$,times,\$10 {{Gb}}/s (1.03 {{Tb}}/s) {{Wavelength Conversion With}} a {{Quasi-Phase-Matched LiNbO}}\$\_3\${{Waveguide}}},}\ }\href {\doibase 10.1109/JSTQE.2006.876150} {\bibfield  {journal} {\bibinfo  {journal} {IEEE Journal of Selected Topics in Quantum Electronics}\ }\textbf {\bibinfo {volume} {12}},\ \bibinfo {pages} {521--528} (\bibinfo {year} {2006})}\BibitemShut {NoStop}%
\bibitem [{\citenamefont {Li}, \citenamefont {Davan{\c c}o},\ and\ \citenamefont {Srinivasan}(2016)}]{liEfficientLownoiseSinglephotonlevel2016}%
  \BibitemOpen
  \bibfield  {author} {\bibinfo {author} {\bibfnamefont {Q.}~\bibnamefont {Li}}, \bibinfo {author} {\bibfnamefont {M.}~\bibnamefont {Davan{\c c}o}}, \ and\ \bibinfo {author} {\bibfnamefont {K.}~\bibnamefont {Srinivasan}},\ }\bibfield  {title} {\enquote {\bibinfo {title} {Efficient and low-noise single-photon-level frequency conversion interfaces using silicon nanophotonics},}\ }\href {\doibase 10.1038/nphoton.2016.64} {\bibfield  {journal} {\bibinfo  {journal} {Nature Photonics}\ }\textbf {\bibinfo {volume} {10}},\ \bibinfo {pages} {406--414} (\bibinfo {year} {2016})}\BibitemShut {NoStop}%
\bibitem [{\citenamefont {Singh}\ \emph {et~al.}(2019)\citenamefont {Singh}, \citenamefont {Li}, \citenamefont {Liu}, \citenamefont {Yu}, \citenamefont {Lu}, \citenamefont {Schneider}, \citenamefont {H{\"o}fling}, \citenamefont {Lawall}, \citenamefont {Verma}, \citenamefont {Mirin}, \citenamefont {Nam}, \citenamefont {Liu},\ and\ \citenamefont {Srinivasan}}]{singhQuantumFrequencyConversion2019}%
  \BibitemOpen
  \bibfield  {author} {\bibinfo {author} {\bibfnamefont {A.}~\bibnamefont {Singh}}, \bibinfo {author} {\bibfnamefont {Q.}~\bibnamefont {Li}}, \bibinfo {author} {\bibfnamefont {S.}~\bibnamefont {Liu}}, \bibinfo {author} {\bibfnamefont {Y.}~\bibnamefont {Yu}}, \bibinfo {author} {\bibfnamefont {X.}~\bibnamefont {Lu}}, \bibinfo {author} {\bibfnamefont {C.}~\bibnamefont {Schneider}}, \bibinfo {author} {\bibfnamefont {S.}~\bibnamefont {H{\"o}fling}}, \bibinfo {author} {\bibfnamefont {J.}~\bibnamefont {Lawall}}, \bibinfo {author} {\bibfnamefont {V.}~\bibnamefont {Verma}}, \bibinfo {author} {\bibfnamefont {R.}~\bibnamefont {Mirin}}, \bibinfo {author} {\bibfnamefont {S.~W.}\ \bibnamefont {Nam}}, \bibinfo {author} {\bibfnamefont {J.}~\bibnamefont {Liu}}, \ and\ \bibinfo {author} {\bibfnamefont {K.}~\bibnamefont {Srinivasan}},\ }\bibfield  {title} {\enquote {\bibinfo {title} {Quantum frequency conversion of a quantum dot single-photon source on a nanophotonic chip},}\ }\href {\doibase 10.1364/OPTICA.6.000563} {\bibfield
   {journal} {\bibinfo  {journal} {Optica}\ }\textbf {\bibinfo {volume} {6}},\ \bibinfo {pages} {563} (\bibinfo {year} {2019})}\BibitemShut {NoStop}%
\bibitem [{\citenamefont {Joshi}\ \emph {et~al.}(2018)\citenamefont {Joshi}, \citenamefont {Farsi}, \citenamefont {Clemmen}, \citenamefont {Ramelow},\ and\ \citenamefont {Gaeta}}]{joshiFrequencyMultiplexingQuasideterministic2018}%
  \BibitemOpen
  \bibfield  {author} {\bibinfo {author} {\bibfnamefont {C.}~\bibnamefont {Joshi}}, \bibinfo {author} {\bibfnamefont {A.}~\bibnamefont {Farsi}}, \bibinfo {author} {\bibfnamefont {S.}~\bibnamefont {Clemmen}}, \bibinfo {author} {\bibfnamefont {S.}~\bibnamefont {Ramelow}}, \ and\ \bibinfo {author} {\bibfnamefont {A.~L.}\ \bibnamefont {Gaeta}},\ }\bibfield  {title} {\enquote {\bibinfo {title} {Frequency multiplexing for quasi-deterministic heralded single-photon sources},}\ }\href {\doibase 10.1038/s41467-018-03254-4} {\bibfield  {journal} {\bibinfo  {journal} {Nature Communications}\ }\textbf {\bibinfo {volume} {9}},\ \bibinfo {pages} {847} (\bibinfo {year} {2018})}\BibitemShut {NoStop}%
\bibitem [{\citenamefont {Umeki}\ \emph {et~al.}(2016)\citenamefont {Umeki}, \citenamefont {Kazama}, \citenamefont {Sano}, \citenamefont {Shibahara}, \citenamefont {Suzuki}, \citenamefont {Abe}, \citenamefont {Takenouchi},\ and\ \citenamefont {Miyamoto}}]{umekiSimultaneousNonlinearityMitigation2016}%
  \BibitemOpen
  \bibfield  {author} {\bibinfo {author} {\bibfnamefont {T.}~\bibnamefont {Umeki}}, \bibinfo {author} {\bibfnamefont {T.}~\bibnamefont {Kazama}}, \bibinfo {author} {\bibfnamefont {A.}~\bibnamefont {Sano}}, \bibinfo {author} {\bibfnamefont {K.}~\bibnamefont {Shibahara}}, \bibinfo {author} {\bibfnamefont {K.}~\bibnamefont {Suzuki}}, \bibinfo {author} {\bibfnamefont {M.}~\bibnamefont {Abe}}, \bibinfo {author} {\bibfnamefont {H.}~\bibnamefont {Takenouchi}}, \ and\ \bibinfo {author} {\bibfnamefont {Y.}~\bibnamefont {Miyamoto}},\ }\bibfield  {title} {\enquote {\bibinfo {title} {Simultaneous nonlinearity mitigation in 92 {\texttimes} 180-{{Gbit}}/s {{PDM-16QAM}} transmission over 3840 km using {{PPLN-based}} guard-band-less optical phase conjugation},}\ }\href {\doibase 10.1364/OE.24.016945} {\bibfield  {journal} {\bibinfo  {journal} {Optics Express}\ }\textbf {\bibinfo {volume} {24}},\ \bibinfo {pages} {16945} (\bibinfo {year} {2016})}\BibitemShut {NoStop}%
\bibitem [{\citenamefont {Ma}\ \emph {et~al.}(2018)\citenamefont {Ma}, \citenamefont {Liang}, \citenamefont {Chen}, \citenamefont {Gao}, \citenamefont {Zheng}, \citenamefont {Xie}, \citenamefont {Liu}, \citenamefont {Zhang},\ and\ \citenamefont {Pan}}]{maUpconversionSinglephotonDetectors2018}%
  \BibitemOpen
  \bibfield  {author} {\bibinfo {author} {\bibfnamefont {F.}~\bibnamefont {Ma}}, \bibinfo {author} {\bibfnamefont {L.-Y.}\ \bibnamefont {Liang}}, \bibinfo {author} {\bibfnamefont {J.-P.}\ \bibnamefont {Chen}}, \bibinfo {author} {\bibfnamefont {Y.}~\bibnamefont {Gao}}, \bibinfo {author} {\bibfnamefont {M.-Y.}\ \bibnamefont {Zheng}}, \bibinfo {author} {\bibfnamefont {X.-P.}\ \bibnamefont {Xie}}, \bibinfo {author} {\bibfnamefont {H.}~\bibnamefont {Liu}}, \bibinfo {author} {\bibfnamefont {Q.}~\bibnamefont {Zhang}}, \ and\ \bibinfo {author} {\bibfnamefont {J.-W.}\ \bibnamefont {Pan}},\ }\bibfield  {title} {\enquote {\bibinfo {title} {Upconversion single-photon detectors based on integrated periodically poled lithium niobate waveguides [{{Invited}}]},}\ }\href {\doibase 10.1364/JOSAB.35.002096} {\bibfield  {journal} {\bibinfo  {journal} {Journal of the Optical Society of America B}\ }\textbf {\bibinfo {volume} {35}},\ \bibinfo {pages} {2096} (\bibinfo {year} {2018})}\BibitemShut {NoStop}%
\bibitem [{\citenamefont {Wang}\ \emph {et~al.}(2023)\citenamefont {Wang}, \citenamefont {Jiao}, \citenamefont {Wang}, \citenamefont {Liu}, \citenamefont {Xie}, \citenamefont {Zheng}, \citenamefont {Zhang},\ and\ \citenamefont {Pan}}]{wangQuantumFrequencyConversion2023a}%
  \BibitemOpen
  \bibfield  {author} {\bibinfo {author} {\bibfnamefont {X.}~\bibnamefont {Wang}}, \bibinfo {author} {\bibfnamefont {X.}~\bibnamefont {Jiao}}, \bibinfo {author} {\bibfnamefont {B.}~\bibnamefont {Wang}}, \bibinfo {author} {\bibfnamefont {Y.}~\bibnamefont {Liu}}, \bibinfo {author} {\bibfnamefont {X.-P.}\ \bibnamefont {Xie}}, \bibinfo {author} {\bibfnamefont {M.-Y.}\ \bibnamefont {Zheng}}, \bibinfo {author} {\bibfnamefont {Q.}~\bibnamefont {Zhang}}, \ and\ \bibinfo {author} {\bibfnamefont {J.-W.}\ \bibnamefont {Pan}},\ }\bibfield  {title} {\enquote {\bibinfo {title} {Quantum frequency conversion and single-photon detection with lithium niobate nanophotonic chips},}\ }\href {\doibase 10.1038/s41534-023-00704-w} {\bibfield  {journal} {\bibinfo  {journal} {npj Quantum Information}\ }\textbf {\bibinfo {volume} {9}},\ \bibinfo {pages} {38} (\bibinfo {year} {2023})}\BibitemShut {NoStop}%
\bibitem [{\citenamefont {Summers}\ \emph {et~al.}(2007)\citenamefont {Summers}, \citenamefont {Masanovic}, \citenamefont {Lal},\ and\ \citenamefont {Blumenthal}}]{summersDesignOperationMonolithically2007}%
  \BibitemOpen
  \bibfield  {author} {\bibinfo {author} {\bibfnamefont {J.~A.}\ \bibnamefont {Summers}}, \bibinfo {author} {\bibfnamefont {M.~L.}\ \bibnamefont {Masanovic}}, \bibinfo {author} {\bibfnamefont {V.}~\bibnamefont {Lal}}, \ and\ \bibinfo {author} {\bibfnamefont {D.~J.}\ \bibnamefont {Blumenthal}},\ }\bibfield  {title} {\enquote {\bibinfo {title} {Design and {{Operation}} of a {{Monolithically Integrated Two-Stage Tunable All-Optical Wavelength Converter}}},}\ }\href {\doibase 10.1109/LPT.2007.902247} {\bibfield  {journal} {\bibinfo  {journal} {IEEE Photonics Technology Letters}\ }\textbf {\bibinfo {volume} {19}},\ \bibinfo {pages} {1248--1250} (\bibinfo {year} {2007})}\BibitemShut {NoStop}%
\bibitem [{\citenamefont {Diez}\ \emph {et~al.}(1997)\citenamefont {Diez}, \citenamefont {Schmidt}, \citenamefont {Ludwig}, \citenamefont {Weber}, \citenamefont {Obermann}, \citenamefont {Kindt}, \citenamefont {Koltchanov},\ and\ \citenamefont {Petermann}}]{diezFourwaveMixingSemiconductor1997}%
  \BibitemOpen
  \bibfield  {author} {\bibinfo {author} {\bibfnamefont {S.}~\bibnamefont {Diez}}, \bibinfo {author} {\bibfnamefont {C.}~\bibnamefont {Schmidt}}, \bibinfo {author} {\bibfnamefont {R.}~\bibnamefont {Ludwig}}, \bibinfo {author} {\bibfnamefont {H.}~\bibnamefont {Weber}}, \bibinfo {author} {\bibfnamefont {K.}~\bibnamefont {Obermann}}, \bibinfo {author} {\bibfnamefont {S.}~\bibnamefont {Kindt}}, \bibinfo {author} {\bibfnamefont {I.}~\bibnamefont {Koltchanov}}, \ and\ \bibinfo {author} {\bibfnamefont {K.}~\bibnamefont {Petermann}},\ }\bibfield  {title} {\enquote {\bibinfo {title} {Four-wave mixing in semiconductor optical amplifiers for frequency conversion and fast optical switching},}\ }\href {\doibase 10.1109/2944.658587} {\bibfield  {journal} {\bibinfo  {journal} {IEEE Journal of Selected Topics in Quantum Electronics}\ }\textbf {\bibinfo {volume} {3}},\ \bibinfo {pages} {1131--1145} (\bibinfo {year} {1997})}\BibitemShut {NoStop}%
\bibitem [{\citenamefont {Lacava}\ \emph {et~al.}(2016)\citenamefont {Lacava}, \citenamefont {Ettabib}, \citenamefont {Cristiani}, \citenamefont {Fedeli}, \citenamefont {Richardson},\ and\ \citenamefont {Petropoulos}}]{lacavaUltraCompactAmorphousSilicon2016}%
  \BibitemOpen
  \bibfield  {author} {\bibinfo {author} {\bibfnamefont {C.}~\bibnamefont {Lacava}}, \bibinfo {author} {\bibfnamefont {M.~A.}\ \bibnamefont {Ettabib}}, \bibinfo {author} {\bibfnamefont {I.}~\bibnamefont {Cristiani}}, \bibinfo {author} {\bibfnamefont {J.~M.}\ \bibnamefont {Fedeli}}, \bibinfo {author} {\bibfnamefont {D.~J.}\ \bibnamefont {Richardson}}, \ and\ \bibinfo {author} {\bibfnamefont {P.}~\bibnamefont {Petropoulos}},\ }\bibfield  {title} {\enquote {\bibinfo {title} {Ultra-{{Compact Amorphous Silicon Waveguide}} for {{Wavelength Conversion}}},}\ }\href {\doibase 10.1109/LPT.2015.2496758} {\bibfield  {journal} {\bibinfo  {journal} {IEEE Photonics Technology Letters}\ }\textbf {\bibinfo {volume} {28}},\ \bibinfo {pages} {410--413} (\bibinfo {year} {2016})}\BibitemShut {NoStop}%
\bibitem [{\citenamefont {Lee}\ \emph {et~al.}(2009)\citenamefont {Lee}, \citenamefont {Biberman}, \citenamefont {{Turner-Foster}}, \citenamefont {Foster}, \citenamefont {Lipson}, \citenamefont {Gaeta},\ and\ \citenamefont {Bergman}}]{leeDemonstrationBroadbandWavelength2009}%
  \BibitemOpen
  \bibfield  {author} {\bibinfo {author} {\bibfnamefont {B.}~\bibnamefont {Lee}}, \bibinfo {author} {\bibfnamefont {A.}~\bibnamefont {Biberman}}, \bibinfo {author} {\bibfnamefont {A.}~\bibnamefont {{Turner-Foster}}}, \bibinfo {author} {\bibfnamefont {M.}~\bibnamefont {Foster}}, \bibinfo {author} {\bibfnamefont {M.}~\bibnamefont {Lipson}}, \bibinfo {author} {\bibfnamefont {A.}~\bibnamefont {Gaeta}}, \ and\ \bibinfo {author} {\bibfnamefont {K.}~\bibnamefont {Bergman}},\ }\bibfield  {title} {\enquote {\bibinfo {title} {Demonstration of {{Broadband Wavelength Conversion}} at 40 {{Gb}}/s in {{Silicon Waveguides}}},}\ }\href {\doibase 10.1109/LPT.2008.2009945} {\bibfield  {journal} {\bibinfo  {journal} {IEEE Photonics Technology Letters}\ }\textbf {\bibinfo {volume} {21}},\ \bibinfo {pages} {182--184} (\bibinfo {year} {2009})}\BibitemShut {NoStop}%
\bibitem [{\citenamefont {Riemensberger}\ \emph {et~al.}(2022)\citenamefont {Riemensberger}, \citenamefont {Kuznetsov}, \citenamefont {Liu}, \citenamefont {He}, \citenamefont {Wang},\ and\ \citenamefont {Kippenberg}}]{riemensbergerPhotonicIntegratedContinuoustravellingwave2022}%
  \BibitemOpen
  \bibfield  {author} {\bibinfo {author} {\bibfnamefont {J.}~\bibnamefont {Riemensberger}}, \bibinfo {author} {\bibfnamefont {N.}~\bibnamefont {Kuznetsov}}, \bibinfo {author} {\bibfnamefont {J.}~\bibnamefont {Liu}}, \bibinfo {author} {\bibfnamefont {J.}~\bibnamefont {He}}, \bibinfo {author} {\bibfnamefont {R.~N.}\ \bibnamefont {Wang}}, \ and\ \bibinfo {author} {\bibfnamefont {T.~J.}\ \bibnamefont {Kippenberg}},\ }\bibfield  {title} {\enquote {\bibinfo {title} {A photonic integrated continuous-travelling-wave parametric amplifier},}\ }\href {\doibase 10.1038/s41586-022-05329-1} {\bibfield  {journal} {\bibinfo  {journal} {Nature}\ }\textbf {\bibinfo {volume} {612}},\ \bibinfo {pages} {56--61} (\bibinfo {year} {2022})}\BibitemShut {NoStop}%
\bibitem [{\citenamefont {Absil}\ \emph {et~al.}(2000)\citenamefont {Absil}, \citenamefont {Hryniewicz}, \citenamefont {Little}, \citenamefont {Cho}, \citenamefont {Wilson}, \citenamefont {Joneckis},\ and\ \citenamefont {Ho}}]{absilWavelengthConversionGaAs2000}%
  \BibitemOpen
  \bibfield  {author} {\bibinfo {author} {\bibfnamefont {P.~P.}\ \bibnamefont {Absil}}, \bibinfo {author} {\bibfnamefont {J.~V.}\ \bibnamefont {Hryniewicz}}, \bibinfo {author} {\bibfnamefont {B.~E.}\ \bibnamefont {Little}}, \bibinfo {author} {\bibfnamefont {P.~S.}\ \bibnamefont {Cho}}, \bibinfo {author} {\bibfnamefont {R.~A.}\ \bibnamefont {Wilson}}, \bibinfo {author} {\bibfnamefont {L.~G.}\ \bibnamefont {Joneckis}}, \ and\ \bibinfo {author} {\bibfnamefont {P.-T.}\ \bibnamefont {Ho}},\ }\bibfield  {title} {\enquote {\bibinfo {title} {Wavelength conversion in {{GaAs}} micro-ring resonators},}\ }\href {\doibase 10.1364/OL.25.000554} {\bibfield  {journal} {\bibinfo  {journal} {Optics Letters}\ }\textbf {\bibinfo {volume} {25}},\ \bibinfo {pages} {554} (\bibinfo {year} {2000})}\BibitemShut {NoStop}%
\bibitem [{\citenamefont {{Haisheng Rong}}\ \emph {et~al.}(2008)\citenamefont {{Haisheng Rong}}, \citenamefont {{Shengbo Xu}}, \citenamefont {Ayotte}, \citenamefont {Cohen}, \citenamefont {Raday},\ and\ \citenamefont {Paniccia}}]{haishengrongSiliconBasedChipscale2008}%
  \BibitemOpen
  \bibfield  {author} {\bibinfo {author} {\bibnamefont {{Haisheng Rong}}}, \bibinfo {author} {\bibnamefont {{Shengbo Xu}}}, \bibinfo {author} {\bibfnamefont {S.}~\bibnamefont {Ayotte}}, \bibinfo {author} {\bibfnamefont {O.}~\bibnamefont {Cohen}}, \bibinfo {author} {\bibfnamefont {O.}~\bibnamefont {Raday}}, \ and\ \bibinfo {author} {\bibfnamefont {M.}~\bibnamefont {Paniccia}},\ }\bibfield  {title} {\enquote {\bibinfo {title} {Silicon based chip-scale nonlinear optical devices: {{Laser}}, amplifier, and wavelength converter},}\ }in\ \href {\doibase 10.1109/LEOSWT.2008.4444373} {\emph {\bibinfo {booktitle} {2008 {{IEEE}}/{{LEOS Winter Topical Meeting Series}}}}}\ (\bibinfo  {publisher} {IEEE},\ \bibinfo {address} {Sorrento, Italy},\ \bibinfo {year} {2008})\ pp.\ \bibinfo {pages} {8--9}\BibitemShut {NoStop}%
\bibitem [{\citenamefont {Kazama}\ \emph {et~al.}(2021)\citenamefont {Kazama}, \citenamefont {Umeki}, \citenamefont {Shimizu}, \citenamefont {Kashiwazaki}, \citenamefont {Enbutsu}, \citenamefont {Kasahara}, \citenamefont {Miyamoto},\ and\ \citenamefont {Watanabe}}]{kazamaOver30dBGain1dB2021b}%
  \BibitemOpen
  \bibfield  {author} {\bibinfo {author} {\bibfnamefont {T.}~\bibnamefont {Kazama}}, \bibinfo {author} {\bibfnamefont {T.}~\bibnamefont {Umeki}}, \bibinfo {author} {\bibfnamefont {S.}~\bibnamefont {Shimizu}}, \bibinfo {author} {\bibfnamefont {T.}~\bibnamefont {Kashiwazaki}}, \bibinfo {author} {\bibfnamefont {K.}~\bibnamefont {Enbutsu}}, \bibinfo {author} {\bibfnamefont {R.}~\bibnamefont {Kasahara}}, \bibinfo {author} {\bibfnamefont {Y.}~\bibnamefont {Miyamoto}}, \ and\ \bibinfo {author} {\bibfnamefont {K.}~\bibnamefont {Watanabe}},\ }\bibfield  {title} {\enquote {\bibinfo {title} {Over-30-{{dB}} gain and 1-{{dB}} noise figure phase-sensitive amplification using a pump-combiner-integrated fiber {{I}}/{{O PPLN}} module},}\ }\href {\doibase 10.1364/OE.434601} {\bibfield  {journal} {\bibinfo  {journal} {Optics Express}\ }\textbf {\bibinfo {volume} {29}},\ \bibinfo {pages} {28824} (\bibinfo {year} {2021})}\BibitemShut {NoStop}%
\bibitem [{\citenamefont {Szabo}\ \emph {et~al.}(2014)\citenamefont {Szabo}, \citenamefont {Puttnam}, \citenamefont {Mazroa}, \citenamefont {Albuquerque}, \citenamefont {Shinada},\ and\ \citenamefont {Wada}}]{szaboNumericalComparisonWDM2014}%
  \BibitemOpen
  \bibfield  {author} {\bibinfo {author} {\bibfnamefont {A.}~\bibnamefont {Szabo}}, \bibinfo {author} {\bibfnamefont {B.~J.}\ \bibnamefont {Puttnam}}, \bibinfo {author} {\bibfnamefont {D.}~\bibnamefont {Mazroa}}, \bibinfo {author} {\bibfnamefont {A.}~\bibnamefont {Albuquerque}}, \bibinfo {author} {\bibfnamefont {S.}~\bibnamefont {Shinada}}, \ and\ \bibinfo {author} {\bibfnamefont {N.}~\bibnamefont {Wada}},\ }\bibfield  {title} {\enquote {\bibinfo {title} {Numerical {{Comparison}} of {{WDM Interchannel Crosstalk}} in {{FOPA-}} and {{PPLN-Based PSAs}}},}\ }\href {\doibase 10.1109/LPT.2014.2327133} {\bibfield  {journal} {\bibinfo  {journal} {IEEE Photonics Technology Letters}\ }\textbf {\bibinfo {volume} {26}},\ \bibinfo {pages} {1503--1506} (\bibinfo {year} {2014})}\BibitemShut {NoStop}%
\bibitem [{\citenamefont {Kazama}\ \emph {et~al.}(2017)\citenamefont {Kazama}, \citenamefont {Umeki}, \citenamefont {Abe}, \citenamefont {Enbutsu}, \citenamefont {Miyamoto},\ and\ \citenamefont {Takenouchi}}]{kazamaLowParametricCrosstalkPhaseSensitiveAmplifier2017c}%
  \BibitemOpen
  \bibfield  {author} {\bibinfo {author} {\bibfnamefont {T.}~\bibnamefont {Kazama}}, \bibinfo {author} {\bibfnamefont {T.}~\bibnamefont {Umeki}}, \bibinfo {author} {\bibfnamefont {M.}~\bibnamefont {Abe}}, \bibinfo {author} {\bibfnamefont {K.}~\bibnamefont {Enbutsu}}, \bibinfo {author} {\bibfnamefont {Y.}~\bibnamefont {Miyamoto}}, \ and\ \bibinfo {author} {\bibfnamefont {H.}~\bibnamefont {Takenouchi}},\ }\bibfield  {title} {\enquote {\bibinfo {title} {Low-{{Parametric-Crosstalk Phase-Sensitive Amplifier}} for {{Guard-Band-Less DWDM Signal Using PPLN Waveguides}}},}\ }\href {\doibase 10.1109/JLT.2016.2603186} {\bibfield  {journal} {\bibinfo  {journal} {Journal of Lightwave Technology}\ }\textbf {\bibinfo {volume} {35}},\ \bibinfo {pages} {755--761} (\bibinfo {year} {2017})}\BibitemShut {NoStop}%
\bibitem [{\citenamefont {Wang}\ \emph {et~al.}(2018)\citenamefont {Wang}, \citenamefont {Langrock}, \citenamefont {Marandi}, \citenamefont {Jankowski}, \citenamefont {Zhang}, \citenamefont {Desiatov}, \citenamefont {Fejer},\ and\ \citenamefont {Lon{\v c}ar}}]{wangUltrahighefficiencyWavelengthConversion2018a}%
  \BibitemOpen
  \bibfield  {author} {\bibinfo {author} {\bibfnamefont {C.}~\bibnamefont {Wang}}, \bibinfo {author} {\bibfnamefont {C.}~\bibnamefont {Langrock}}, \bibinfo {author} {\bibfnamefont {A.}~\bibnamefont {Marandi}}, \bibinfo {author} {\bibfnamefont {M.}~\bibnamefont {Jankowski}}, \bibinfo {author} {\bibfnamefont {M.}~\bibnamefont {Zhang}}, \bibinfo {author} {\bibfnamefont {B.}~\bibnamefont {Desiatov}}, \bibinfo {author} {\bibfnamefont {M.~M.}\ \bibnamefont {Fejer}}, \ and\ \bibinfo {author} {\bibfnamefont {M.}~\bibnamefont {Lon{\v c}ar}},\ }\bibfield  {title} {\enquote {\bibinfo {title} {Ultrahigh-efficiency wavelength conversion in nanophotonic periodically poled lithium niobate waveguides},}\ }\href {\doibase 10.1364/OPTICA.5.001438} {\bibfield  {journal} {\bibinfo  {journal} {Optica}\ }\textbf {\bibinfo {volume} {5}},\ \bibinfo {pages} {1438} (\bibinfo {year} {2018})}\BibitemShut {NoStop}%
\bibitem [{\citenamefont {Chen}\ \emph {et~al.}(2024{\natexlab{a}})\citenamefont {Chen}, \citenamefont {Briggs}, \citenamefont {Cui}, \citenamefont {Zhang}, \citenamefont {Shah},\ and\ \citenamefont {Fan}}]{chenAdaptedPolingBreak2024}%
  \BibitemOpen
  \bibfield  {author} {\bibinfo {author} {\bibfnamefont {P.-K.}\ \bibnamefont {Chen}}, \bibinfo {author} {\bibfnamefont {I.}~\bibnamefont {Briggs}}, \bibinfo {author} {\bibfnamefont {C.}~\bibnamefont {Cui}}, \bibinfo {author} {\bibfnamefont {L.}~\bibnamefont {Zhang}}, \bibinfo {author} {\bibfnamefont {M.}~\bibnamefont {Shah}}, \ and\ \bibinfo {author} {\bibfnamefont {L.}~\bibnamefont {Fan}},\ }\bibfield  {title} {\enquote {\bibinfo {title} {Adapted poling to break the nonlinear efficiency limit in nanophotonic lithium niobate waveguides},}\ }\href {\doibase 10.1038/s41565-023-01525-w} {\bibfield  {journal} {\bibinfo  {journal} {Nature Nanotechnology}\ }\textbf {\bibinfo {volume} {19}},\ \bibinfo {pages} {44--50} (\bibinfo {year} {2024}{\natexlab{a}})}\BibitemShut {NoStop}%
\bibitem [{\citenamefont {He}\ \emph {et~al.}(2024)\citenamefont {He}, \citenamefont {Liu}, \citenamefont {Lin}, \citenamefont {Chen},\ and\ \citenamefont {Ma}}]{heEfficientSecondHarmonicGeneration2024}%
  \BibitemOpen
  \bibfield  {author} {\bibinfo {author} {\bibfnamefont {J.}~\bibnamefont {He}}, \bibinfo {author} {\bibfnamefont {L.}~\bibnamefont {Liu}}, \bibinfo {author} {\bibfnamefont {M.}~\bibnamefont {Lin}}, \bibinfo {author} {\bibfnamefont {H.}~\bibnamefont {Chen}}, \ and\ \bibinfo {author} {\bibfnamefont {F.}~\bibnamefont {Ma}},\ }\bibfield  {title} {\enquote {\bibinfo {title} {Efficient {{Second-Harmonic Generation}} in {{Adapted-Width Waveguides Based}} on {{Periodically Poled Thin-Film Lithium Niobate}}},}\ }\href {\doibase 10.3390/mi15091145} {\bibfield  {journal} {\bibinfo  {journal} {Micromachines}\ }\textbf {\bibinfo {volume} {15}},\ \bibinfo {pages} {1145} (\bibinfo {year} {2024})}\BibitemShut {NoStop}%
\bibitem [{\citenamefont {Jankowski}\ \emph {et~al.}(2020)\citenamefont {Jankowski}, \citenamefont {Langrock}, \citenamefont {Desiatov}, \citenamefont {Marandi}, \citenamefont {Wang}, \citenamefont {Zhang}, \citenamefont {Phillips}, \citenamefont {Lon{\v c}ar},\ and\ \citenamefont {Fejer}}]{jankowskiUltrabroadbandNonlinearOptics2020a}%
  \BibitemOpen
  \bibfield  {author} {\bibinfo {author} {\bibfnamefont {M.}~\bibnamefont {Jankowski}}, \bibinfo {author} {\bibfnamefont {C.}~\bibnamefont {Langrock}}, \bibinfo {author} {\bibfnamefont {B.}~\bibnamefont {Desiatov}}, \bibinfo {author} {\bibfnamefont {A.}~\bibnamefont {Marandi}}, \bibinfo {author} {\bibfnamefont {C.}~\bibnamefont {Wang}}, \bibinfo {author} {\bibfnamefont {M.}~\bibnamefont {Zhang}}, \bibinfo {author} {\bibfnamefont {C.~R.}\ \bibnamefont {Phillips}}, \bibinfo {author} {\bibfnamefont {M.}~\bibnamefont {Lon{\v c}ar}}, \ and\ \bibinfo {author} {\bibfnamefont {M.~M.}\ \bibnamefont {Fejer}},\ }\bibfield  {title} {\enquote {\bibinfo {title} {Ultrabroadband nonlinear optics in nanophotonic periodically poled lithium niobate waveguides},}\ }\href {\doibase 10.1364/OPTICA.7.000040} {\bibfield  {journal} {\bibinfo  {journal} {Optica}\ }\textbf {\bibinfo {volume} {7}},\ \bibinfo {pages} {40} (\bibinfo {year} {2020})}\BibitemShut {NoStop}%
\bibitem [{\citenamefont {Li}\ \emph {et~al.}(2024)\citenamefont {Li}, \citenamefont {Li}, \citenamefont {Wang}, \citenamefont {Chen}, \citenamefont {Ma}, \citenamefont {Zhang}, \citenamefont {Sun},\ and\ \citenamefont {Wang}}]{liAdvancingLargescaleThinfilm2024}%
  \BibitemOpen
  \bibfield  {author} {\bibinfo {author} {\bibfnamefont {X.}~\bibnamefont {Li}}, \bibinfo {author} {\bibfnamefont {H.}~\bibnamefont {Li}}, \bibinfo {author} {\bibfnamefont {Z.}~\bibnamefont {Wang}}, \bibinfo {author} {\bibfnamefont {Z.}~\bibnamefont {Chen}}, \bibinfo {author} {\bibfnamefont {F.}~\bibnamefont {Ma}}, \bibinfo {author} {\bibfnamefont {K.}~\bibnamefont {Zhang}}, \bibinfo {author} {\bibfnamefont {W.}~\bibnamefont {Sun}}, \ and\ \bibinfo {author} {\bibfnamefont {C.}~\bibnamefont {Wang}},\ }\bibfield  {title} {\enquote {\bibinfo {title} {Advancing large-scale thin-film {{PPLN}} nonlinear photonics with segmented tunable micro-heaters},}\ }\href {\doibase 10.1364/PRJ.516180} {\bibfield  {journal} {\bibinfo  {journal} {Photonics Research}\ }\textbf {\bibinfo {volume} {12}},\ \bibinfo {pages} {1703} (\bibinfo {year} {2024})}\BibitemShut {NoStop}%
\bibitem [{\citenamefont {Stokowski}\ \emph {et~al.}(2024)\citenamefont {Stokowski}, \citenamefont {Dean}, \citenamefont {Hwang}, \citenamefont {Park}, \citenamefont {Celik}, \citenamefont {McKenna}, \citenamefont {Jankowski}, \citenamefont {Langrock}, \citenamefont {Ansari}, \citenamefont {Fejer},\ and\ \citenamefont {{Safavi-Naeini}}}]{stokowskiIntegratedFrequencymodulatedOptical2024b}%
  \BibitemOpen
  \bibfield  {author} {\bibinfo {author} {\bibfnamefont {H.~S.}\ \bibnamefont {Stokowski}}, \bibinfo {author} {\bibfnamefont {D.~J.}\ \bibnamefont {Dean}}, \bibinfo {author} {\bibfnamefont {A.~Y.}\ \bibnamefont {Hwang}}, \bibinfo {author} {\bibfnamefont {T.}~\bibnamefont {Park}}, \bibinfo {author} {\bibfnamefont {O.~T.}\ \bibnamefont {Celik}}, \bibinfo {author} {\bibfnamefont {T.~P.}\ \bibnamefont {McKenna}}, \bibinfo {author} {\bibfnamefont {M.}~\bibnamefont {Jankowski}}, \bibinfo {author} {\bibfnamefont {C.}~\bibnamefont {Langrock}}, \bibinfo {author} {\bibfnamefont {V.}~\bibnamefont {Ansari}}, \bibinfo {author} {\bibfnamefont {M.~M.}\ \bibnamefont {Fejer}}, \ and\ \bibinfo {author} {\bibfnamefont {A.~H.}\ \bibnamefont {{Safavi-Naeini}}},\ }\bibfield  {title} {\enquote {\bibinfo {title} {Integrated frequency-modulated optical parametric oscillator},}\ }\href {\doibase 10.1038/s41586-024-07071-2} {\bibfield  {journal} {\bibinfo  {journal} {Nature}\ }\textbf {\bibinfo {volume} {627}},\ \bibinfo {pages}
  {95--100} (\bibinfo {year} {2024})}\BibitemShut {NoStop}%
\bibitem [{\citenamefont {Ledezma}\ \emph {et~al.}(2023)\citenamefont {Ledezma}, \citenamefont {Roy}, \citenamefont {Costa}, \citenamefont {Sekine}, \citenamefont {Gray}, \citenamefont {Guo}, \citenamefont {Nehra}, \citenamefont {Briggs},\ and\ \citenamefont {Marandi}}]{ledezmaOctavespanningTunableInfrared2023d}%
  \BibitemOpen
  \bibfield  {author} {\bibinfo {author} {\bibfnamefont {L.}~\bibnamefont {Ledezma}}, \bibinfo {author} {\bibfnamefont {A.}~\bibnamefont {Roy}}, \bibinfo {author} {\bibfnamefont {L.}~\bibnamefont {Costa}}, \bibinfo {author} {\bibfnamefont {R.}~\bibnamefont {Sekine}}, \bibinfo {author} {\bibfnamefont {R.}~\bibnamefont {Gray}}, \bibinfo {author} {\bibfnamefont {Q.}~\bibnamefont {Guo}}, \bibinfo {author} {\bibfnamefont {R.}~\bibnamefont {Nehra}}, \bibinfo {author} {\bibfnamefont {R.~M.}\ \bibnamefont {Briggs}}, \ and\ \bibinfo {author} {\bibfnamefont {A.}~\bibnamefont {Marandi}},\ }\bibfield  {title} {\enquote {\bibinfo {title} {Octave-spanning tunable infrared parametric oscillators in nanophotonics},}\ }\href {\doibase 10.1126/sciadv.adf9711} {\bibfield  {journal} {\bibinfo  {journal} {Science Advances}\ }\textbf {\bibinfo {volume} {9}},\ \bibinfo {pages} {eadf9711} (\bibinfo {year} {2023})}\BibitemShut {NoStop}%
\bibitem [{\citenamefont {Hwang}\ \emph {et~al.}(2023)\citenamefont {Hwang}, \citenamefont {Stokowski}, \citenamefont {Park}, \citenamefont {Jankowski}, \citenamefont {McKenna}, \citenamefont {Langrock}, \citenamefont {Mishra}, \citenamefont {Ansari}, \citenamefont {Fejer},\ and\ \citenamefont {{Safavi-Naeini}}}]{hwangMidinfraredSpectroscopyBroadly2023f}%
  \BibitemOpen
  \bibfield  {author} {\bibinfo {author} {\bibfnamefont {A.~Y.}\ \bibnamefont {Hwang}}, \bibinfo {author} {\bibfnamefont {H.~S.}\ \bibnamefont {Stokowski}}, \bibinfo {author} {\bibfnamefont {T.}~\bibnamefont {Park}}, \bibinfo {author} {\bibfnamefont {M.}~\bibnamefont {Jankowski}}, \bibinfo {author} {\bibfnamefont {T.~P.}\ \bibnamefont {McKenna}}, \bibinfo {author} {\bibfnamefont {C.}~\bibnamefont {Langrock}}, \bibinfo {author} {\bibfnamefont {J.}~\bibnamefont {Mishra}}, \bibinfo {author} {\bibfnamefont {V.}~\bibnamefont {Ansari}}, \bibinfo {author} {\bibfnamefont {M.~M.}\ \bibnamefont {Fejer}}, \ and\ \bibinfo {author} {\bibfnamefont {A.~H.}\ \bibnamefont {{Safavi-Naeini}}},\ }\bibfield  {title} {\enquote {\bibinfo {title} {Mid-infrared spectroscopy with a broadly tunable thin-film lithium niobate optical parametric oscillator},}\ }\href {\doibase 10.1364/OPTICA.502487} {\bibfield  {journal} {\bibinfo  {journal} {Optica}\ }\textbf {\bibinfo {volume} {10}},\ \bibinfo {pages} {1535} (\bibinfo {year}
  {2023})}\BibitemShut {NoStop}%
\bibitem [{\citenamefont {Li}\ \emph {et~al.}(2022)\citenamefont {Li}, \citenamefont {Chang}, \citenamefont {Wu}, \citenamefont {Staffa}, \citenamefont {Ling}, \citenamefont {Javid}, \citenamefont {Xue}, \citenamefont {He}, \citenamefont {{Lopez-rios}}, \citenamefont {Morin}, \citenamefont {Wang}, \citenamefont {Shen}, \citenamefont {Zeng}, \citenamefont {Zhu}, \citenamefont {Vahala}, \citenamefont {Bowers},\ and\ \citenamefont {Lin}}]{liIntegratedPockelsLaser2022e}%
  \BibitemOpen
  \bibfield  {author} {\bibinfo {author} {\bibfnamefont {M.}~\bibnamefont {Li}}, \bibinfo {author} {\bibfnamefont {L.}~\bibnamefont {Chang}}, \bibinfo {author} {\bibfnamefont {L.}~\bibnamefont {Wu}}, \bibinfo {author} {\bibfnamefont {J.}~\bibnamefont {Staffa}}, \bibinfo {author} {\bibfnamefont {J.}~\bibnamefont {Ling}}, \bibinfo {author} {\bibfnamefont {U.~A.}\ \bibnamefont {Javid}}, \bibinfo {author} {\bibfnamefont {S.}~\bibnamefont {Xue}}, \bibinfo {author} {\bibfnamefont {Y.}~\bibnamefont {He}}, \bibinfo {author} {\bibfnamefont {R.}~\bibnamefont {{Lopez-rios}}}, \bibinfo {author} {\bibfnamefont {T.~J.}\ \bibnamefont {Morin}}, \bibinfo {author} {\bibfnamefont {H.}~\bibnamefont {Wang}}, \bibinfo {author} {\bibfnamefont {B.}~\bibnamefont {Shen}}, \bibinfo {author} {\bibfnamefont {S.}~\bibnamefont {Zeng}}, \bibinfo {author} {\bibfnamefont {L.}~\bibnamefont {Zhu}}, \bibinfo {author} {\bibfnamefont {K.~J.}\ \bibnamefont {Vahala}}, \bibinfo {author} {\bibfnamefont {J.~E.}\ \bibnamefont {Bowers}}, \ and\ \bibinfo
  {author} {\bibfnamefont {Q.}~\bibnamefont {Lin}},\ }\bibfield  {title} {\enquote {\bibinfo {title} {Integrated {{Pockels}} laser},}\ }\href {\doibase 10.1038/s41467-022-33101-6} {\bibfield  {journal} {\bibinfo  {journal} {Nature Communications}\ }\textbf {\bibinfo {volume} {13}},\ \bibinfo {pages} {5344} (\bibinfo {year} {2022})}\BibitemShut {NoStop}%
\bibitem [{\citenamefont {Wei}\ \emph {et~al.}(2024)\citenamefont {Wei}, \citenamefont {Cheng}, \citenamefont {Wu}, \citenamefont {Zeng}, \citenamefont {Tang},\ and\ \citenamefont {Xia}}]{weiEfficientBroadbandAllOptical2024}%
  \BibitemOpen
  \bibfield  {author} {\bibinfo {author} {\bibfnamefont {J.}~\bibnamefont {Wei}}, \bibinfo {author} {\bibfnamefont {C.}~\bibnamefont {Cheng}}, \bibinfo {author} {\bibfnamefont {Y.}~\bibnamefont {Wu}}, \bibinfo {author} {\bibfnamefont {C.}~\bibnamefont {Zeng}}, \bibinfo {author} {\bibfnamefont {M.}~\bibnamefont {Tang}}, \ and\ \bibinfo {author} {\bibfnamefont {J.}~\bibnamefont {Xia}},\ }\bibfield  {title} {\enquote {\bibinfo {title} {Efficient and {{Broadband All-Optical Wavelength Conversion Between C}} and {{O Bands}} in {{Nanophotonic Lithium Niobate Waveguides}}},}\ }\href {\doibase 10.1109/JLT.2024.3406579} {\bibfield  {journal} {\bibinfo  {journal} {Journal of Lightwave Technology}\ }\textbf {\bibinfo {volume} {42}},\ \bibinfo {pages} {5966--5973} (\bibinfo {year} {2024})}\BibitemShut {NoStop}%
\bibitem [{\citenamefont {Wei}\ \emph {et~al.}(2022)\citenamefont {Wei}, \citenamefont {Hu}, \citenamefont {Zhang}, \citenamefont {Li}, \citenamefont {Wu}, \citenamefont {Zeng}, \citenamefont {Tang},\ and\ \citenamefont {Xia}}]{weiAllopticalWavelengthConversion2022a}%
  \BibitemOpen
  \bibfield  {author} {\bibinfo {author} {\bibfnamefont {J.}~\bibnamefont {Wei}}, \bibinfo {author} {\bibfnamefont {Z.}~\bibnamefont {Hu}}, \bibinfo {author} {\bibfnamefont {M.}~\bibnamefont {Zhang}}, \bibinfo {author} {\bibfnamefont {P.}~\bibnamefont {Li}}, \bibinfo {author} {\bibfnamefont {Y.}~\bibnamefont {Wu}}, \bibinfo {author} {\bibfnamefont {C.}~\bibnamefont {Zeng}}, \bibinfo {author} {\bibfnamefont {M.}~\bibnamefont {Tang}}, \ and\ \bibinfo {author} {\bibfnamefont {J.}~\bibnamefont {Xia}},\ }\bibfield  {title} {\enquote {\bibinfo {title} {All-optical wavelength conversion of a 92-{{Gb}}/s 16-{{QAM}} signal within the {{C-band}} in a single thin-film {{PPLN}} waveguide},}\ }\href {\doibase 10.1364/OE.465382} {\bibfield  {journal} {\bibinfo  {journal} {Optics Express}\ }\textbf {\bibinfo {volume} {30}},\ \bibinfo {pages} {30564} (\bibinfo {year} {2022})}\BibitemShut {NoStop}%
\bibitem [{\citenamefont {Sun}, \citenamefont {Liu},\ and\ \citenamefont {Yariv}(2009)}]{sunAdiabaticityCriterionShortest2009}%
  \BibitemOpen
  \bibfield  {author} {\bibinfo {author} {\bibfnamefont {X.}~\bibnamefont {Sun}}, \bibinfo {author} {\bibfnamefont {H.-C.}\ \bibnamefont {Liu}}, \ and\ \bibinfo {author} {\bibfnamefont {A.}~\bibnamefont {Yariv}},\ }\bibfield  {title} {\enquote {\bibinfo {title} {Adiabaticity criterion and the shortest adiabatic mode transformer in a coupled-waveguide system},}\ }\href {\doibase 10.1364/OL.34.000280} {\bibfield  {journal} {\bibinfo  {journal} {Optics Letters}\ }\textbf {\bibinfo {volume} {34}},\ \bibinfo {pages} {280} (\bibinfo {year} {2009})}\BibitemShut {NoStop}%
\bibitem [{\citenamefont {Zhu}\ \emph {et~al.}(2024)\citenamefont {Zhu}, \citenamefont {Hu}, \citenamefont {Lu}, \citenamefont {Warner}, \citenamefont {Li}, \citenamefont {Song}, \citenamefont {Magalh{\~a}es}, \citenamefont {{Shams-Ansari}}, \citenamefont {Cordaro}, \citenamefont {Sinclair},\ and\ \citenamefont {Lon{\v c}ar}}]{zhuTwentynineMillionIntrinsic2024}%
  \BibitemOpen
  \bibfield  {author} {\bibinfo {author} {\bibfnamefont {X.}~\bibnamefont {Zhu}}, \bibinfo {author} {\bibfnamefont {Y.}~\bibnamefont {Hu}}, \bibinfo {author} {\bibfnamefont {S.}~\bibnamefont {Lu}}, \bibinfo {author} {\bibfnamefont {H.~K.}\ \bibnamefont {Warner}}, \bibinfo {author} {\bibfnamefont {X.}~\bibnamefont {Li}}, \bibinfo {author} {\bibfnamefont {Y.}~\bibnamefont {Song}}, \bibinfo {author} {\bibfnamefont {L.}~\bibnamefont {Magalh{\~a}es}}, \bibinfo {author} {\bibfnamefont {A.}~\bibnamefont {{Shams-Ansari}}}, \bibinfo {author} {\bibfnamefont {A.}~\bibnamefont {Cordaro}}, \bibinfo {author} {\bibfnamefont {N.}~\bibnamefont {Sinclair}}, \ and\ \bibinfo {author} {\bibfnamefont {M.}~\bibnamefont {Lon{\v c}ar}},\ }\bibfield  {title} {\enquote {\bibinfo {title} {Twenty-nine million intrinsic {{{\emph{Q}}}} -factor monolithic microresonators on thin-film lithium niobate},}\ }\href {\doibase 10.1364/PRJ.521172} {\bibfield  {journal} {\bibinfo  {journal} {Photonics Research}\ }\textbf {\bibinfo {volume} {12}},\
  \bibinfo {pages} {A63} (\bibinfo {year} {2024})}\BibitemShut {NoStop}%
\bibitem [{\citenamefont {Zhao}\ \emph {et~al.}(2020)\citenamefont {Zhao}, \citenamefont {R{\"u}sing}, \citenamefont {Javid}, \citenamefont {Ling}, \citenamefont {Li}, \citenamefont {Lin},\ and\ \citenamefont {Mookherjea}}]{zhaoShallowetchedThinfilmLithium2020h}%
  \BibitemOpen
  \bibfield  {author} {\bibinfo {author} {\bibfnamefont {J.}~\bibnamefont {Zhao}}, \bibinfo {author} {\bibfnamefont {M.}~\bibnamefont {R{\"u}sing}}, \bibinfo {author} {\bibfnamefont {U.~A.}\ \bibnamefont {Javid}}, \bibinfo {author} {\bibfnamefont {J.}~\bibnamefont {Ling}}, \bibinfo {author} {\bibfnamefont {M.}~\bibnamefont {Li}}, \bibinfo {author} {\bibfnamefont {Q.}~\bibnamefont {Lin}}, \ and\ \bibinfo {author} {\bibfnamefont {S.}~\bibnamefont {Mookherjea}},\ }\bibfield  {title} {\enquote {\bibinfo {title} {Shallow-etched thin-film lithium niobate waveguides for highly-efficient second-harmonic generation},}\ }\href {\doibase 10.1364/OE.395545} {\bibfield  {journal} {\bibinfo  {journal} {Optics Express}\ }\textbf {\bibinfo {volume} {28}},\ \bibinfo {pages} {19669} (\bibinfo {year} {2020})}\BibitemShut {NoStop}%
\bibitem [{\citenamefont {Chang}\ \emph {et~al.}(2016)\citenamefont {Chang}, \citenamefont {Li}, \citenamefont {Volet}, \citenamefont {Wang}, \citenamefont {Peters},\ and\ \citenamefont {Bowers}}]{changThinFilmWavelength2016d}%
  \BibitemOpen
  \bibfield  {author} {\bibinfo {author} {\bibfnamefont {L.}~\bibnamefont {Chang}}, \bibinfo {author} {\bibfnamefont {Y.}~\bibnamefont {Li}}, \bibinfo {author} {\bibfnamefont {N.}~\bibnamefont {Volet}}, \bibinfo {author} {\bibfnamefont {L.}~\bibnamefont {Wang}}, \bibinfo {author} {\bibfnamefont {J.}~\bibnamefont {Peters}}, \ and\ \bibinfo {author} {\bibfnamefont {J.~E.}\ \bibnamefont {Bowers}},\ }\bibfield  {title} {\enquote {\bibinfo {title} {Thin film wavelength converters for photonic integrated circuits},}\ }\href {\doibase 10.1364/OPTICA.3.000531} {\bibfield  {journal} {\bibinfo  {journal} {Optica}\ }\textbf {\bibinfo {volume} {3}},\ \bibinfo {pages} {531} (\bibinfo {year} {2016})}\BibitemShut {NoStop}%
\bibitem [{\citenamefont {Zhao}, \citenamefont {R{\"u}sing},\ and\ \citenamefont {Mookherjea}(2019)}]{zhaoOpticalDiagnosticMethods2019b}%
  \BibitemOpen
  \bibfield  {author} {\bibinfo {author} {\bibfnamefont {J.}~\bibnamefont {Zhao}}, \bibinfo {author} {\bibfnamefont {M.}~\bibnamefont {R{\"u}sing}}, \ and\ \bibinfo {author} {\bibfnamefont {S.}~\bibnamefont {Mookherjea}},\ }\bibfield  {title} {\enquote {\bibinfo {title} {Optical diagnostic methods for monitoring the poling of thin-film lithium niobate waveguides},}\ }\href {\doibase 10.1364/OE.27.012025} {\bibfield  {journal} {\bibinfo  {journal} {Optics Express}\ }\textbf {\bibinfo {volume} {27}},\ \bibinfo {pages} {12025} (\bibinfo {year} {2019})}\BibitemShut {NoStop}%
\bibitem [{\citenamefont {Nagy}, \citenamefont {Prabhakar},\ and\ \citenamefont {Reano}(2020)}]{nagySituTemporalPeriodic2020}%
  \BibitemOpen
  \bibfield  {author} {\bibinfo {author} {\bibfnamefont {J.~T.}\ \bibnamefont {Nagy}}, \bibinfo {author} {\bibfnamefont {K.}~\bibnamefont {Prabhakar}}, \ and\ \bibinfo {author} {\bibfnamefont {R.~M.}\ \bibnamefont {Reano}},\ }\bibfield  {title} {\enquote {\bibinfo {title} {In {{Situ Temporal Periodic Poling}} of {{Lithium Niobate Thin Films}}},}\ }in\ \href {\doibase 10.1364/CLEO_SI.2020.SW3F.3} {\emph {\bibinfo {booktitle} {Conference on {{Lasers}} and {{Electro-Optics}}}}}\ (\bibinfo  {publisher} {Optica Publishing Group},\ \bibinfo {address} {Washington, DC},\ \bibinfo {year} {2020})\ p.\ \bibinfo {pages} {SW3F.3}\BibitemShut {NoStop}%
\bibitem [{\citenamefont {Chen}\ \emph {et~al.}(2024{\natexlab{b}})\citenamefont {Chen}, \citenamefont {Wang}, \citenamefont {Jia}, \citenamefont {Tian}, \citenamefont {Tang}, \citenamefont {Zhu}, \citenamefont {Gu}, \citenamefont {Zhao}, \citenamefont {Wang}, \citenamefont {Ye}, \citenamefont {Tang}, \citenamefont {Zhang}, \citenamefont {Yan}, \citenamefont {Qian}, \citenamefont {Jin}, \citenamefont {Wang}, \citenamefont {Zhu},\ and\ \citenamefont {Xie}}]{chenHighgainOpticalParametric2024}%
  \BibitemOpen
  \bibfield  {author} {\bibinfo {author} {\bibfnamefont {M.}~\bibnamefont {Chen}}, \bibinfo {author} {\bibfnamefont {C.}~\bibnamefont {Wang}}, \bibinfo {author} {\bibfnamefont {K.}~\bibnamefont {Jia}}, \bibinfo {author} {\bibfnamefont {X.-H.}\ \bibnamefont {Tian}}, \bibinfo {author} {\bibfnamefont {J.}~\bibnamefont {Tang}}, \bibinfo {author} {\bibfnamefont {C.}~\bibnamefont {Zhu}}, \bibinfo {author} {\bibfnamefont {X.}~\bibnamefont {Gu}}, \bibinfo {author} {\bibfnamefont {Z.}~\bibnamefont {Zhao}}, \bibinfo {author} {\bibfnamefont {Z.}~\bibnamefont {Wang}}, \bibinfo {author} {\bibfnamefont {Z.}~\bibnamefont {Ye}}, \bibinfo {author} {\bibfnamefont {J.}~\bibnamefont {Tang}}, \bibinfo {author} {\bibfnamefont {Y.}~\bibnamefont {Zhang}}, \bibinfo {author} {\bibfnamefont {Z.}~\bibnamefont {Yan}}, \bibinfo {author} {\bibfnamefont {G.}~\bibnamefont {Qian}}, \bibinfo {author} {\bibfnamefont {B.}~\bibnamefont {Jin}}, \bibinfo {author} {\bibfnamefont {Z.}~\bibnamefont {Wang}}, \bibinfo {author} {\bibfnamefont {S.-N.}\
  \bibnamefont {Zhu}}, \ and\ \bibinfo {author} {\bibfnamefont {Z.}~\bibnamefont {Xie}},\ }\href {\doibase 10.48550/arXiv.2411.10721} {\enquote {\bibinfo {title} {High-gain optical parametric amplification with continuous-wave pump using domain-engineered thin film lithium niobate waveguide},}\ } (\bibinfo {year} {2024}{\natexlab{b}}),\ \Eprint {http://arxiv.org/abs/2411.10721} {arXiv:2411.10721 [physics]} \BibitemShut {NoStop}%
\bibitem [{\citenamefont {Maeder}\ \emph {et~al.}(2022)\citenamefont {Maeder}, \citenamefont {Kaufmann}, \citenamefont {Pohl}, \citenamefont {Kellner},\ and\ \citenamefont {Grange}}]{maederHighbandwidthThermoopticPhase2022}%
  \BibitemOpen
  \bibfield  {author} {\bibinfo {author} {\bibfnamefont {A.}~\bibnamefont {Maeder}}, \bibinfo {author} {\bibfnamefont {F.}~\bibnamefont {Kaufmann}}, \bibinfo {author} {\bibfnamefont {D.}~\bibnamefont {Pohl}}, \bibinfo {author} {\bibfnamefont {J.}~\bibnamefont {Kellner}}, \ and\ \bibinfo {author} {\bibfnamefont {R.}~\bibnamefont {Grange}},\ }\bibfield  {title} {\enquote {\bibinfo {title} {High-bandwidth thermo-optic phase shifters for lithium niobate-on-insulator photonic integrated circuits},}\ }\href {\doibase 10.1364/OL.469358} {\bibfield  {journal} {\bibinfo  {journal} {Optics Letters}\ }\textbf {\bibinfo {volume} {47}},\ \bibinfo {pages} {4375} (\bibinfo {year} {2022})}\BibitemShut {NoStop}%
\end{thebibliography}%

\end{document}